\documentclass{nature}

\usepackage{graphicx}
\makeatletter
\let\saved@includegraphics\includegraphics
\AtBeginDocument{\let\includegraphics\saved@includegraphics}

\makeatother

\usepackage{multirow}
\usepackage{subfigure}
\usepackage{mathtools}
\usepackage[colorlinks=true,linkcolor=blue,citecolor=black,urlcolor=blue]{hyperref}
\usepackage[dvipsnames]{xcolor}
\usepackage{soul}
\usepackage{authblk}
\usepackage[labelfont=bf]{caption}
\usepackage[figurename=Figure]{caption}
\usepackage{siunitx}
\usepackage{booktabs}

\newcommand{\ud}{\mathrm{d}}
\newcommand{\iu}{\mathrm{i}}

\newcommand{\IM}{\mathrm{Im}}

\newcommand{\VEC}[1]{\mathbf{#1}}

\begin{document}

\title{Spin-orbit torques and their associated effective fields from gigahertz to terahertz}

\author[1*]{Filipe S. M. Guimar\~aes}
\author{Juba Bouaziz}
\author{Manuel dos Santos Dias}
\author{Samir Lounis}
\affil{Peter Gr\"unberg Institut and Institute for Advanced Simulation, Forschungszentrum J\"ulich \& JARA, J\"ulich D-52425, Germany}
\affil[*]{f.guimaraes@fz-juelich.de}

\maketitle

\begin{abstract}

Terahertz spintronics offers the prospect of devices which are both faster and more energy-efficient.
A promising route to achieve this goal is to exploit current-induced spin-orbit torques.
However, the high-frequency properties of these quantities remain unexplored both experimentally and theoretically, within a realistic material-specific approach.
Here we investigate the dynamical transverse components of the torques and uncover contributions longitudinal to the magnetic moment capable of changing its magnitude. 
We show that, while the torques can be drastically altered in the dynamical regime, the effective magnetic fields that accompany them present a frequency-independent behaviour, ranging from the static limit up to the terahertz domain --- including the ferromagnetic resonance of the system.
The outcomes of this work point to new ways to control magnetic units in next-generation spintronic devices.

\end{abstract}

\section*{Introduction}

A great number of prospected technological applications are based on the effective manipulation of the magnetisation at ultrafast speeds~\cite{Bigot:2013ih,Walowski:2016bt}.
This involves the application of torques caused either by spin-polarized currents (spin-transfer torques)~\cite{Ralph:2008kj} or by spin currents and accumulations induced by the spin-orbit interaction (SOI)~\cite{Manchon:2018vg,Ramaswamy:2018bc}.
The latter are the so-called spin-orbit torques (SOTs), which provide an efficient way to switch the magnetisation in systems containing heavy metals~\cite{Miron:2011gd,Liu:2012co} or topological insulators~\cite{Wang:2017hp,Han:2017bu,DC:2018js} at room temperature.
SOTs are also active in laser-excited ferromagnetic structures in the terahertz (THz) range~\cite{SeifertT:2016kc}, which also comprises the frequencies of excitations of antiferromagnetic materials~\cite{Bhattacharjee:2018es}.
In fact, SOTs were used to manipulate N\'eel vectors between different states of an antiferromagnet~\cite{Olejnik:2018bk}.
The manipulation of magnetic states by the current-induced torques provides novel capabilities for future technology --- including memristors, building blocks for neuromorphic computing~\cite{Torrejon:2017hj,Zhang:2019cx}.

For ultrafast applications, knowledge of the dynamical behaviour of the SOT is paramount.
The SOT is usually quantified experimentally by its angular dependence with respect to the magnetisation direction~\cite{Garello:2013fa,Chen:2018iq,Safranski:2018kl}, while its behaviour with the frequency can be measured via spin-torque ferromagnetic resonance~\cite{Liu:2011fp,Fang:2011ch}.
However, the frequency dependence is obtained indirectly through a rectified voltage generated by the mixing signals of an alternating current (AC) and the precessing magnetisation~\cite{Liu:2011fp,Zhou:2016bn,Safranski:2018kl}.
Second-harmonic techniques are also used to measure the effective magnetic fields that the SOT induce on the magnetisation, but up to now only in the quasi-static regime~\cite{Garello:2013fa,Ghosh:2017ib}.

From the theoretical side, calculations of the SOTs based on realistic electronic structures were set forth~\cite{Freimuth:2014kq,Freimuth:2015fq,Belashchenko:2019et}, with a few  predictions for the angular dependencies~\cite{Haney:2013bq,Mahfouzi:2018jg,Belashchenko:2019et}.
They are unobliging with the frequency dependence since they are restricted to static perturbations.
The dynamics of the magnetisation is then commonly described by phenomenological equations~\cite{Manchon:2009gp,Haney:2013bp,Legrand:2015fz,Mahfouzi:2018jg,Baek:2018hp,Bhattacharjee:2018es}.
Thus, even though the great potential of the spin-orbit torque lies in the dynamics it induces, there is a big gap in the knowledge of its time or frequency dependence.

Conventionally, the SOTs are decomposed into a field-like and a damping-like component, in analogy to the equation of motion for a magnetic moment subjected to external magnetic fields~\cite{Gilbert:2004gx}.
While the field-like torque drives the magnetic moment into precessional motion around the spin accumulation direction, the damping-like torque points to/away from it, depending on the direction of the applied field. 
Due to the symmetry of these components with the magnetisation direction (odd and even, respectively), it is expected that large damping-like torques are required for an effective manipulation of the magnetic moment~\cite{Chen:2018iq,Zhu:2019dw,Zhu:2019fa}, albeit field-like ones can also cause the switching in the presence of assisting magnetic fields~\cite{Wang:2012dt}.
In reality, the multi-parameter magnetisation dynamics can follow rather intricate paths, and particular combinations of field-like and damping-like torques can be used to induce deterministic switching of the magnetisation without assisting fields~\cite{Legrand:2015fz}.

In this work, we investigate the spin-orbit torques starting from the microscopic theory of the magnetisation dynamics in realistic materials.
This theory incorporates the collective spin excitations in the presence of spin-orbit interaction, inherently taking into account the intrinsic interfacial and bulk spin-orbit-related mechanisms, such as the inverse spin galvanic effect and the spin Hall effect, respectively, as well as their reciprocal phenomena~\cite{Manchon:2015hr,Guimaraes:2017kk}.
We fully unveil the frequency and angular dependency of the SOTs in ferromagnetic/heavy metal bilayer structures subjected to an external AC electric field.
The SOTs are highly frequency dependent, changing more than one order of magnitude with respect to the static values. 
We also uncover a counter-intuitive component of the torque capable of altering the magnetisation length.
Its amplitude is found to be one order of magnitude smaller than the usual SOTs, but also shares a rich and complex frequency dependence.
Surprisingly, the effective magnetic fields that provide equivalent description of the magnetisation dynamics weakly depend on the frequency for the investigated bilayers. 
These findings enlighten how to use the torques for the manipulation of the magnetisation using time-dependent electric currents.

\section*{Results}

\subsection{Theory of spin-orbit torques and effective magnetic fields}

In an atomistic picture, the central quantity is the spin magnetic moment $\VEC{M}_i(t)$ belonging to atom $i$, which is obtained microscopically from the quantum-mechanical expectation value of the spin angular momentum operator.
Its dynamics are governed by the spin continuity equation
\begin{equation}
\label{eq:mov}
\frac{\ud \VEC{M}_i(t)}{\ud t} = -\VEC{I}^\text{S}_i(t)+ \boldsymbol{\tau}^\text{ext}_i(t)+ \boldsymbol{\tau}^\text{SOI}_i(t) \quad .
\end{equation}
It includes the effects of the spin currents $\VEC{I}^\text{S}_i = \sum_j \VEC{I}^\text{S}_{ij}(t)$, where $ \VEC{I}^\text{S}_{ij}(t)$ denotes the spin currents flowing from atom $i$ to neighboring atoms $j$, the Zeeman torque due to the external magnetic field, $ \boldsymbol{\tau}^\text{ext}_i(t)= \VEC{M}_i(t)\times \VEC{B}^\text{ext}$, and the local spin-orbit torque $ \boldsymbol{\tau}^\text{SOI}_i(t)= \lambda_i\,\langle\VEC{S}_i\times \VEC{L}_i\rangle(t)$, which is due to the atomic spin-orbit interaction with strength $\lambda_i$ and couples spin and orbital angular momentum.
It is important to remark that the expectation value of the local spin orbit torque operator cannot be obtained from the simple cross product of the local spin and orbital moments.
The terms $\VEC{I}^\text{S}_i$ and $\boldsymbol{\tau}^\text{SOI}_i$ represent transfer of angular momentum out of the spin degree of freedom of atom $i$ and, therefore, can give rise to spin relaxation effects.
The left-hand side of Eq.~\eqref{eq:mov} is zero in the ground state of the system, determining the equilibrium orientation of the spin moment as $\VEC{M}_i = M_i\,\hat{\VEC{m}}_i$ ($M_i$ being its magnitude and $\hat{\VEC{m}}_i$ its direction).
We note that $\VEC{M}_i$ is one order of magnitude larger than the orbital magnetic moment, so it serves as a good proxy for the total magnetisation dynamics.

\begin{figure*}[bt!]
  \centering
  \includegraphics[width=0.8\columnwidth]{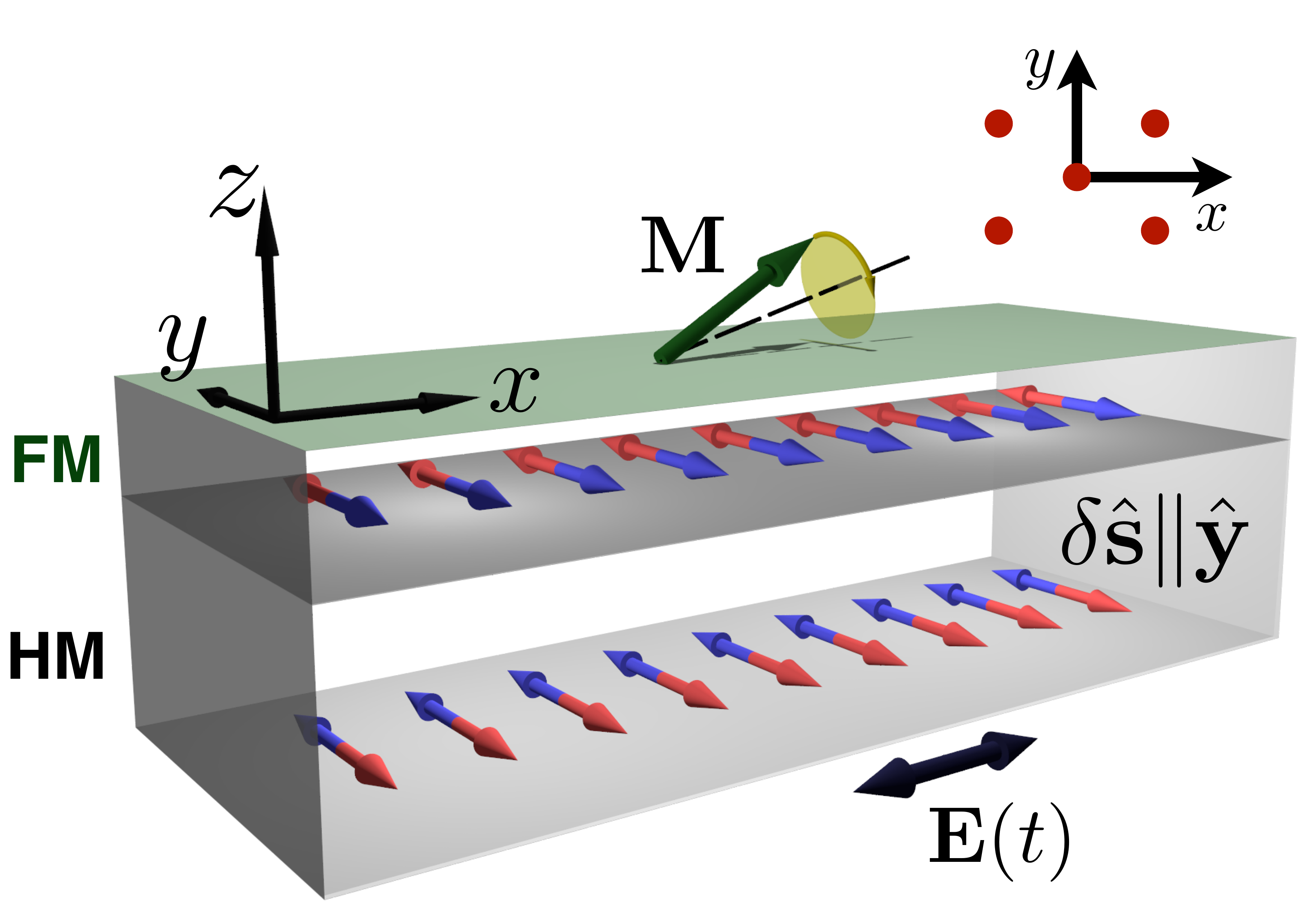}
  \caption{Diagram of the investigated ferromagnetic (FM) / heavy metal (HM) bilayers. A uniform and oscillatory electric field $\VEC{E}(t)$ is applied in the $x$ direction, generating a spin accumulation along the $y$ direction.
  Its interaction with the FM layer drives the magnetic moment $\VEC{M}$ into precessional motion.
  On the top right, a scheme of the $C_{2v}$-symmetric W(110) surface.
\label{fig:diagram}}
\end{figure*}
We investigate bilayers consisting of a ferromagnetic layer (FM) deposited on a heavy metal (HM) substrate. 
The setup is depicted in Fig.~\ref{fig:diagram}. 
The main focus of this work is on the $C_{2v}$-symmetric Fe/W(110), for which the spin moment has an in-plane easy axis along the long axis of the lattice ($x$-axis, as indicated in Fig.~\ref{fig:diagram}).
Results for Co/Pt(001) bilayer (see, e.g., Refs.~\citenum{Miron:2011gd,Garello:2013fa,Mahfouzi:2018jg}), for which the easy axis is out-of-plane, are given in the Supplementary Figs. 1\,--\,4.
The external magnetic field $\VEC{B}^\text{ext}$ is applied to the whole sample and is used to reorient the magnetic moment of the system.

As shown in Fig.~\ref{fig:diagram}, we consider the application of a spatially uniform AC electric field with frequency $\omega$ along the in-plane $x$-direction, $\VEC{E}(t) = E_0 \cos(\omega t)\,\hat{\VEC{x}}$.

This induces a spin current $\VEC{I}^\text{S}_i$ flowing along the $\hat{\VEC{z}}$ direction and an oscillatory spin accumulation $\delta \langle\VEC{S}_i\rangle(t)\propto\hat{\VEC{z}}\times\VEC{E}(t)$ due to bulk and interfacial effects~\cite{Manchon:2018vg}, which is thus polarized along $\hat{\VEC{y}}$ and builds up at the interface between the FM and the HM.
The total spin accumulation may have other contributions that are dependent on the direction of the magnetization of the FM.
This can be taken into account through the dependence of the field-like and damping-like contributions to the spin-orbit torque on the orientation of the magnetization of the ferromagnet.
The electric field also modifies the orbital angular momentum $\delta\langle \VEC{L}_i\rangle(t)$ in the ferromagnet.
All these contributions then interact with the magnetic moment $\VEC{M}_i$ of the FM layer via the total spin-orbit torques $\boldsymbol{\tau}^\text{SOT}_i(t) = -\VEC{I}^\text{S}_i(t) + \boldsymbol{\tau}^\text{SOI}_i(t)$, driving the system out of equilibrium according to Eq.~\eqref{eq:mov}.

The magnetisation dynamics induced by the external electric field is described by Eq.~\eqref{eq:mov}.
The SOT $\boldsymbol{\tau}^\text{SOT}_i$ is obtained within linear response theory in terms of the vector potential $\VEC{A}(t) = -\frac{E_0}{\omega}\sin(\omega t)\,\VEC{\hat{x}}$ (as detailed in the Supplementary Note 1).
The harmonic perturbation gives rise to the frequency-dependent components ($\alpha=x,y,z$) of the spin-orbit torque
\begin{equation}\label{eq:freqdep}
\begin{split}
\tau^\text{SOT}_{i,\alpha}(t)&= E_0\left[\tau^\text{in-phase}_{i,\alpha}(\omega)\cos(\omega t) - \tau^\text{out-phase}_{i,\alpha}(\omega)\sin(\omega t) \right]
\\ &= E_0|\tau^\text{SOT}_{i,\alpha}(\omega)|\cos\Big(\omega t-\varphi^\text{SOT}_{i,\alpha}(\omega)\Big)\ ,
\end{split}
\end{equation}
which includes contributions that are in-phase and out-of-phase with the external electric field.
The amplitude of the oscillation is given by $|\tau^\text{SOT}_{i,\alpha}(\omega)| = \sqrt{\tau^\text{in-phase}_{i,\alpha}(\omega)^2 + \tau^\text{out-phase}_{i,\alpha}(\omega)^2}$, and the dephasing may also be expressed in terms of the phase of the response function, $\varphi^\text{SOT}_{i,\alpha}(\omega)$.
The coefficients $\tau^\text{in-phase}_{i,\alpha}(\omega)$, $\tau^\text{out-phase}_{i,\alpha}(\omega)$ and $|\tau^\text{SOT}_{i,\alpha}(\omega)|$, weighted by the electric field $E_0$, are also known as \textit{torkance}. 
In the static limit, $\tau^\text{in-phase}_{i,\alpha}$ was obtained in Refs.~\citenum{Freimuth:2014kq,Belashchenko:2019et}.
We consider the SOTs in the local frame of reference of the atomic magnetic moments and decompose them as
\begin{equation}\label{eq:localframe}
\boldsymbol{\tau}^\text{SOT}_i(\omega) = \tau_{i,\text{FL}}(\omega)\,\hat{\VEC{m}}_i\times\delta \hat{\VEC{s}}
+ \tau_{i,\text{DL}}(\omega)\,\hat{\VEC{m}}_i\times(\hat{\VEC{m}}_i\times\delta\hat{\VEC{s}}) + \tau_{i,\|}(\omega)\,\hat{\VEC{m}}_i\ ,
\end{equation}
where $\delta \hat{\VEC{s}}$ represents the direction of the spin accumulation, $\tau_{i,\text{FL}}(\omega)$ and $\tau_{i,\text{DL}}(\omega)$ are the field-like and damping-like (or antidamping-like) SOT strengths~\cite{Freimuth:2014kq}, respectively, and $\tau_{i,\|}(\omega)$ is the strength of the longitudinal component, i.e., along the magnetisation direction.

Spin-orbit torque experiments are often interpreted in terms of an \textit{effective magnetic field}~\cite{Garello:2013fa,Ghosh:2017ib}: the fictitious magnetic field that would cause the same perturbation on the magnetic moments as the real applied electric field.
The real dynamical response of the system is given by
\begin{align} \label{eq:spincurrent}
\delta \VEC{M}_i(\omega) =& \sum_{j}\underline{\Xi}_{ij}(\omega) \VEC{A}_j(\omega)\ ,
\end{align}
where $\underline{\Xi}(\omega)$ represent the magnetic-charge current response.
Then, the obtained perturbation of the magnetic moment $\delta \VEC{M}_i(\omega)$ is inserted in the definition of the magnetic susceptibility $\underline{\chi}(\omega)$,~\cite{Freimuth:2015fq}
\begin{align} \label{eq:spinspin}
\delta \VEC{M}_i(\omega) =& \sum_{j}\underline{\chi}_{ij}(\omega) \VEC{B}_{j}(\omega)\ .
\end{align}
From the solution of Eq.~\eqref{eq:spinspin}, the effective magnetic field caused by the harmonic electric field applied to the system can be cast in the form of Eq.~\eqref{eq:freqdep} as
\begin{align}\label{eq:beff}
\VEC{B}^\text{eff}_i(t) &= \sum_{jk}\IM\left\{[\underline{\chi}(\omega) ]^{-1}_{ij} \underline{\Xi}_{jk}(\omega) \frac{E_0 e^{-\iu\omega t}}{\omega}\VEC{\hat{x}}\right\}\\ \label{eq:beff2}
&= E_0\left[\VEC{B}^\text{in-phase}_i(\omega)\cos(\omega t) - \VEC{B}^\text{out-phase}_i(\omega)\sin(\omega t) \right]\ ,
\end{align}
where the in-phase and out-of-phase coefficients are obtained from the imaginary and real parts of $\VEC{B}_i(\omega) = \frac{1}{\omega}\sum_{jk}[\underline{\chi}(\omega) ]^{-1}_{ij} \underline{\Xi}_{jk}(\omega)\,\VEC{\hat{x}}$, respectively.
The spin-orbit interaction couples the spin and charge degrees of freedom, such that $\underline{\Xi}$ and $\underline{\chi}$ must be written in the complete charge and spin basis for each atom, $\{\sigma^0, \sigma^x,\sigma^y,\sigma^z\} $, where $\sigma^0$ is the identity matrix and the three others are the Pauli matrices.
Eqs.~\eqref{eq:spincurrent} and \eqref{eq:spinspin} use only their relevant parts.
Note that the direction of $\VEC{B}_i(\omega)$ is determined by the structure of these matrices.

In the next sections, these two central quantities --- SOT and effective magnetic field --- are discussed within the framework of linear response theory with respect to the amplitude of the external electric field, $E_0$. 
More details on our approach, based on a multi-orbital tight-binding hamiltonian that captures the electronic structure obtained from a density functional theory calculation, can be found in the Methods section and in the Supplementary Note 1.

\subsection{Spin-orbit torques on Fe/W(110)}

\begin{figure*}[ht!]
  \centering
  \includegraphics[width=1.0\textwidth]{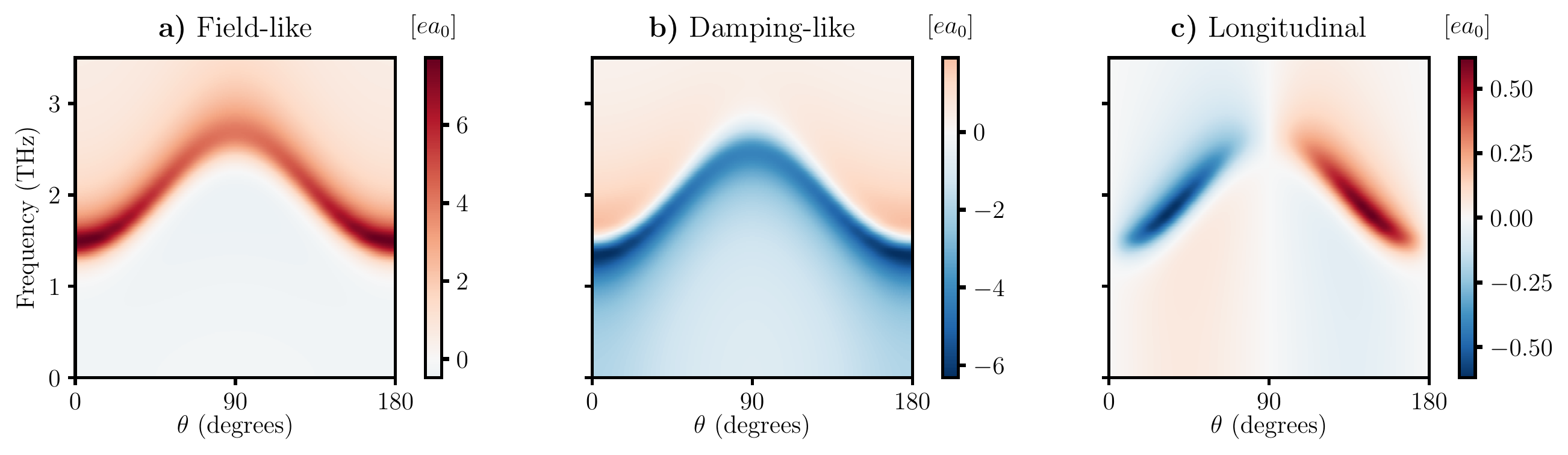}
  \caption{Frequency and angular dependencies of the dynamical spin-orbit torques. Components of the spin-orbit torque $\boldsymbol{\tau}^\text{SOT}$ in the local frame of reference that are in phase with the oscillatory electric field, when the magnetic moment of the ferromagnetic layer in the Fe/W(110) bilayer is rotated within the $zx$ plane by an external magnetic field.
  $z$ is the normal to the film surface and the electric field is applied along $x$, with $\theta$ the angle between the magnetic moment and the surface normal. 
  The torques are obtained in units of $ea_0 =\SI{8.48e-30}{\coulomb\metre}$, where $e$ is the electronic charge and $a_0$ is the Bohr radius.
  The torque vector was decomposed into the local frame of reference given by: (a) a field-like component along $\hat{\VEC{m}}_\text{Fe}\times\delta \hat{\VEC{s}}$; (b) a damping-like component along $\hat{\VEC{m}}_\text{Fe}\times(\hat{\VEC{m}}_\text{Fe}\times\delta \hat{\VEC{s}})$; and (c) a longitudinal contribution along $\hat{\VEC{m}}_\text{Fe}$.
\label{fig:sot_inphase}}
\end{figure*}
We start by examining the SOT components for an Fe/W(110) bilayer subjected to an external AC electric field.
Fig.~\ref{fig:sot_inphase} shows the frequency and angular dependencies of the in-phase torque $\boldsymbol{\tau}^\text{in-phase}_i(\omega)$ acting on the magnetic moment of the Fe layer, since it is the relevant one in the quasi-static regime (out-of-phase $\boldsymbol{\tau}^\text{out-phase}_{i}(\omega)$ and amplitude $|\boldsymbol{\tau}^\text{SOT}_{i}(\omega)|$ are shown in Supplementary Figures 5\,--\,6).
Fig.~\ref{fig:sot_inphase}(a,b) display the projection of the transverse components into the local frame given in Eq.~\eqref{eq:localframe}, while Fig.~\ref{fig:sot_inphase}c depicts the longitudinal component.
The large in-plane anisotropy of a monolayer of Fe on W(110) shifts the resonance position to very high frequencies when $\VEC{M}_\text{Fe}\,\|\,\hat{\VEC{x}}$.
A magnetic field larger than the anisotropy field is used to reorient the magnetic moment within the $zx$ plane (the rotation along $zy$ and $xy$ planes can be found in the Supplementary Figure 7).
When the magnetisation is aligned with the spin accumulation generated by the electric field, i.e., $\VEC{M}_\text{Fe}\,\|\,\delta \hat{\VEC{s}}\,\|\,\hat{\VEC{y}}$, the transverse components of the torque vanish.
As an important remark, the projection of the transverse components into the local frame already takes into account the angular dependency of the torques to lowest order in the magnetisation direction.
Nevertheless, these components of the SOT displayed in Fig.~\ref{fig:sot_inphase} still vary when the magnetisation is rotated.
This is a signature of the higher order terms in $\hat{\VEC{m}}_i$, which substantially alter the angular dependency of the SOTs, as found experimentally in Ref.~\citenum{Garello:2013fa}.

We also note that the projection of $\boldsymbol{\tau}^\text{SOT}_i$ into the local frame of reference results in a non-vanishing longitudinal component (i.e., along the magnetisation direction $\hat{\VEC{m}}_i$),
even when $\VEC{M}_\text{Fe}\,\|\,\delta \VEC{s}$ (as seen in the cuts depicted in Fig.~\ref{fig:sot_cut_inphase} below).
This induces changes in the magnitude of $\VEC{M}_\text{Fe}$ and is also present in the static limit.
When the magnetisation is not along high symmetry directions, this contribution is enhanced due to the presence of spin currents originating from the misalignment of the magnetic moments among the different layers~\cite{Bruno:2005hy,Ruckriegel:2017eg}.
The amplitude of the longitudinal component is approximately one order of magnitude smaller than the transverse ones.
This is attributed to the weaker longitudinal magnetic responses when compared to the transverse ones, as their excitation energy is located in the eV range, which is settled by the energy cost of modifying the length of a magnetic moments~\cite{Kim:1973dd,IbanezAzpiroz:2017gc}.
Interestingly, we observe resonant features in the longitudinal components originating from the transversal excitations, since these responses are coupled due to the spin-orbit interaction.

All components of the torque have a striking variation in the vicinity of the resonance.
Not only there are regions of large increase of the magnitude of the torque, but they also change sign.
The angular dependency is highly influenced by the excitation frequency owing to the resonance of the magnetisation.
In Fig.~\ref{fig:sot_cut_inphase}, the angular dependencies of the torques for two distinct frequencies are shown: the static limit $\omega=0$ (solid lines) and $\omega_\text{res}^{z}=\SI{1.5}{\tera\hertz}$ (dashed lines), which is the resonance frequency for $\VEC{M}_\text{Fe}\parallel \hat{\VEC{z}}$ ($\theta=0$).
While the former presents slightly slanted sinusoidal oscillations (due to the low symmetry of the $C_{2v}$ lattice characterizing the investigated bilayer), the latter is eminently non-trivial due to the complex dependence of the resonance frequency on the magnetisation direction.
Moreover, for the selected frequency $\omega_\text{res}^{z}$, large enhancements of the components are obtained when the magnetisation is rotated in the $zx$ (green curves) and $zy$ (red curves) planes.
\begin{figure*}[ht!]
  \centering
  \includegraphics[width=1.0\textwidth]{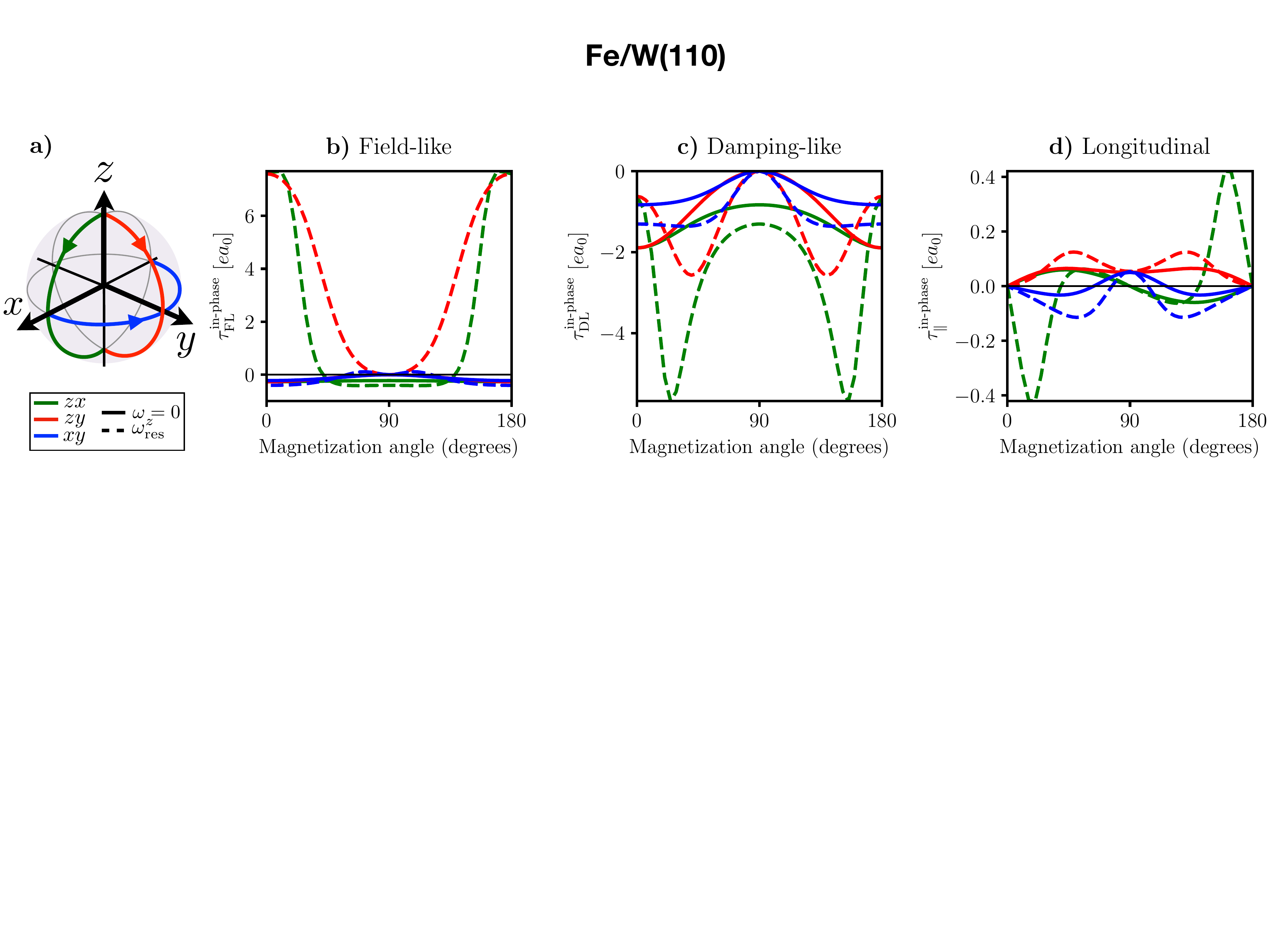}
  \caption{Angular dependencies of spin-orbit torques in Fe/W(110). Components of the spin-orbit torques $\boldsymbol{\tau}^\text{SOT}$ that are in-phase with the oscillatory electric field for the magnetic moment along the directions indicated in (a): $z\rightarrow x$ (green), $z\rightarrow y$ (red) and $x\rightarrow y$ (blue). The torques are decomposed in (b) field-like, (c) damping-like and (d) longitudinal components for two distinct frequencies: static limit $\omega=0$ (solid lines) and $\omega_\text{res}^z=\SI{1.5}{\tera\hertz}$ (dashed lines). The torques are given in units of $ea_0 =\SI{8.48e-30}{\coulomb\metre}$, where $e$ is the electronic charge and $a_0$ is the Bohr radius.}
\label{fig:sot_cut_inphase} 
\end{figure*}

\begin{figure*}[ht!]
  \centering
  \includegraphics[width=1.0\textwidth]{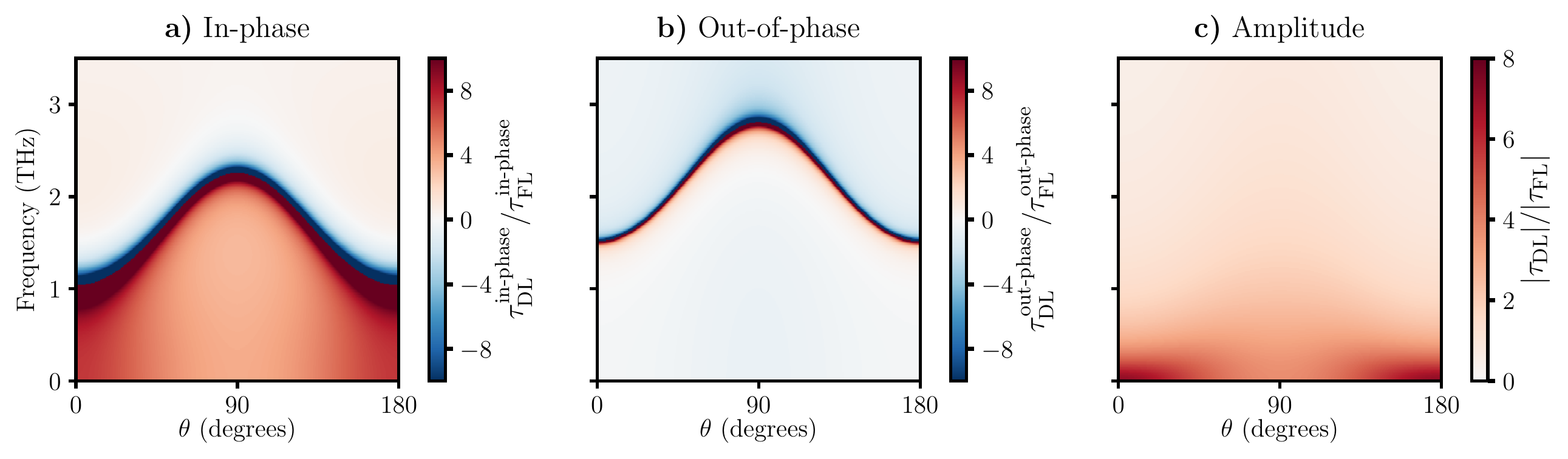}
  \caption{Ratios between damping-like and field-like components of the spin-orbit torques. Frequency and angular dependencies of the ratios between the damping-like, $\tau_{\text{DL}}(\omega)$, and field-like, $\tau_{\text{FL}}(\omega)$, components of the spin-orbit torques: (a) in-phase and (b) out-of-phase contributions, and (c) amplitudes.
  The magnetic moment is rotated in the $zx$ plane, $z$ is the normal to the film surface and the electric field is applied along $x$.
\label{fig:sot_ratio}}
\end{figure*}

The capability of manipulating the torque can be unveiled by calculating the ratio between the damping-like and field-like components of the SOT.
We depict these ratios for the in-phase (Fig.~\ref{fig:sot_ratio}a), out-of-phase (Fig.~\ref{fig:sot_ratio}b) and amplitude (Fig.~\ref{fig:sot_ratio}c) of the torques.
Large values are obtained for the in-phase and out-of-phase ratios whenever the field-like component of the torques crosses zero.
The ratio of the amplitudes displayed in Fig.~\ref{fig:sot_ratio}c does not present any feature close to the resonance and shows that the damping-like component is more important for low frequencies.
The behaviour illustrated in Fig.~\ref{fig:sot_ratio} demonstrates the possibility of tuning the torques by varying the frequency of the electric field or the temporal characteristics of the pulses.

Although the SOTs are directly responsible for the dynamics of the magnetic moment, they can also be quantified in terms of effective magnetic fields acting on $\VEC{M}_\text{Fe}$.
In the next section, we investigate the frequency and angular dependencies of these fields.

\subsection{Effective fields on Fe/W(110)}
\begin{figure*}[ht!]
  \centering
  \includegraphics[width=1.0\textwidth]{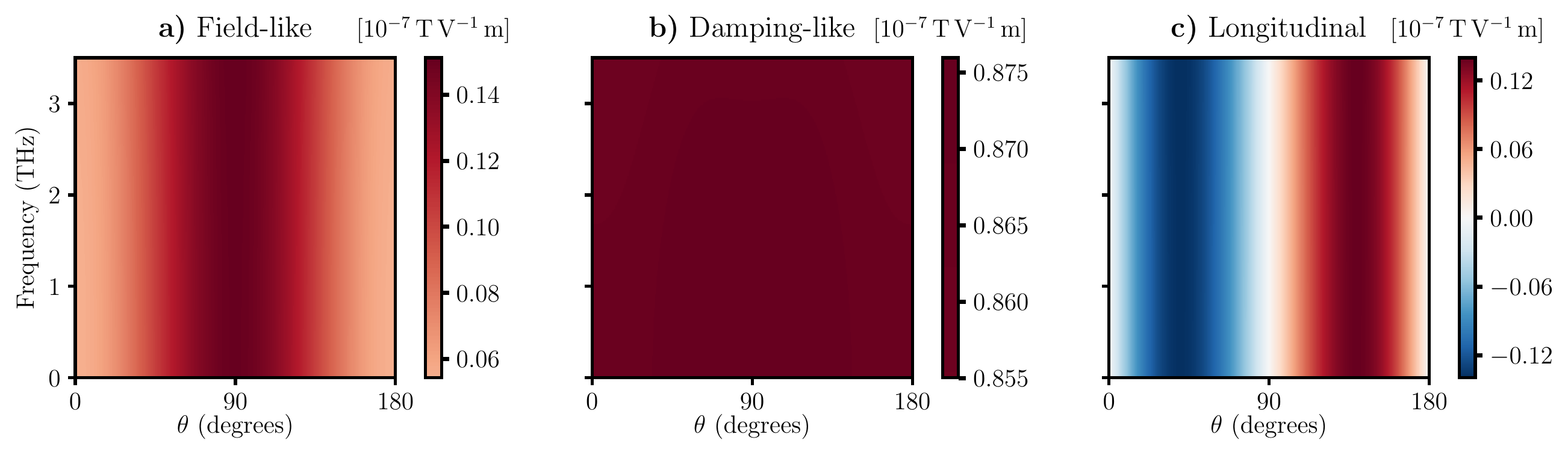}
  \caption{Frequency and angular dependencies of the effective magnetic fields. (a) Field-like, (b) Damping-like and (c) Longitudinal components of the effective field $\VEC{B}^\text{eff}_i(\omega)$, given by Eq.~\eqref{eq:beff}, in the local frame of reference that are in phase with the oscillatory electric field, when the magnetic moment of the ferromagnetic layer in the Fe/W(110) bilayer is rotated within the $zx$ plane by an external magnetic field.
\label{fig:beff_inphase}}
\end{figure*}
We use Eq.~\eqref{eq:beff} to calculate the the effective field in the local frame of reference.
Fig.~\ref{fig:beff_inphase} displays its components as a function of the frequency and magnetisation direction (rotated within the $zx$ plane).
Surprisingly, despite a strong frequency dependence of the torques and the magnetisation, the effective fields remain almost unaltered by the frequency sweep.
This behaviour is a direct consequence of the result obtained in Eq.~\eqref{eq:beff}.
Indeed, the effective field consists of a product of the inverse magnetic susceptibility with the magnetic-charge current one.
The factor $(1+\underline{\chi}_0(\omega) \underline{U})^{-1}$ involving the non-interacting susceptibility $\underline{\chi}_0(\omega)$ and the effective Coulomb interaction $\underline{U}$ that renormalizes both responses~\cite{Costa2010,dosSantosDias:2015bh,Guimaraes:2015fl} then cancels identically (see Supplementary Note 2).
Physically, this lack of structure in the frequency range relevant for spin dynamics shows that the intrinsic timescale for the charge-to-spin conversion is much faster than the one for the spin dynamics.
The angular variation of the transverse projections into the local frame are once more due to the presence of higher order terms in the magnetisation (similarly as discussed for the SOT).
Note that these high order terms are important for the field-like component of the effective field (see Fig.~\ref{fig:beff_inphase}a), but negligible for the damping-like one (Fig.~\ref{fig:beff_inphase}b).
The latter is reflected in respectively small fitting coefficients, as shown in the Supplementary Note 3.
The longitudinal part is found to be of the same order of magnitude as the transverse ones.
However, the smaller responses in this direction lead to weak longitudinal SOTs (as depicted in Fig.~\ref{fig:sot_inphase}c).

\section*{Discussion}

In this paper, we focused on the investigation of the frequency and angular dependencies of the spin-orbit torques and the accompanying effective magnetic fields for Fe/W(110) and Co/Pt(001) bilayers relying on a unified approach that incorporates both magnetic and charge responses on equal footing.
Within this framework, we revealed the presence of non-vanishing longitudinal dynamics capable of altering the magnetisation length.
While the spin-orbit torques exhibit a strong frequency dependency at the vicinity of the ferromagnetic resonance of the system, this is not so for the effective fields.
The angular dependencies of both the SOTs and the effective magnetic fields present important higher order contributions in terms of the magnetisation direction.
Their signature is observed upon rotation of the magnetisation.

Most of the existing literature addressing spin-orbit torques follows the approach described in Ref.~\citenum{Manchon:2009gp}, which is based on the following assumptions: The charge carriers and the ferromagnetic unit are considered separately, interacting with each other via an $sd$-like coupling, and the static limit ($\omega=0$) of the spin continuity equation is taken.
Then, the SOT is obtained from the torque between the magnetic moment of the ferromagnet and the spin moment of the charge carriers.
A similar approach was also employed to obtain the SOT as the cross product between the magnetic moment and the exchange correlation field~\cite{Freimuth:2014kq,Belashchenko:2019et}.
However, for finite frequencies, the dynamics of the non-equilibrium magnetisation must be taken into account.
In addition, when the magnetic moment and the carriers are treated on equal footing --- as in the approach we use here ---, this kind of torque vanishes identically (see Supplementary Note 1).
Moreover, the contrasting behaviour of the computed torques and effective fields with respect to the frequency evidences that the relation $\VEC{B}^\text{eff} = \boldsymbol{\tau}^\text{SOT}\times\hat{\VEC{m}}$ cannot be simply generalized by inserting the corresponding frequency-dependent quantities obtained in linear response.
As a matter of fact, they must be computed in a consistent manner using the form provided in Eq.~\eqref{eq:beff}.
It follows that the appropriate form of the SOT for atomistic spin dynamics simulations is given by the torque between the magnetic moment and the field in Eq.~\eqref{eq:beff2}, i.e., $\hat{\VEC{m}}(t)\times\VEC{B}^\text{eff}(t)$.
This is consistent with the form of the torque which is known from the static limit~\cite{Garello:2013fa,Avci:2014fj,Belashchenko:2019et} and provides a microscopic justification to its use in the time-dependent case.

The effective fields are measured experimentally in the quasi-static regime (few to hundreds of Hz) using, for example, the second harmonic technique~\cite{Garello:2013fa,Ghosh:2017ib}.
As we found them to be weakly frequency dependent, their quasi-static values can then be used to interpret terahertz dynamics.
However, our microscopic theory summarized by Eq.~\eqref{eq:beff} reveals that, in extreme cases where the electronic structure supports strong interband transitions in the terahertz range, the effective fields are expected to acquire a non-trivial frequency dependence.
Such an electronic structure scenario may occur in systems exhibiting strong spin-orbit interaction, as in Tl/Si(111) bilayers~\cite{LafuenteBartolome:2017fz}.

Our theory has a natural synergy with ultrafast experiments, where the magnetic state of the system is manipulated with laser pulses~\cite{Garello:2014bi,Zhang:2016hz,Olejnik:2018bk}.
They can be used to take advantage of the large variations of the SOTs with the frequency by adjusting the ratio between the field-like and the damping-like components of the torques.
In practice, this control can be achieved by applying two pulses to excite the magnetisation.
The delay between them determines if the response to the second pulse is dominated by the in-phase or the out-of-phase component.
This protocol is complementary to previous proposals based on the variation of the amplitude and length of the current pulses~\cite{Miron:2011gd,Garello:2014bi,Olejnik:2015bc} as a mean for writing and reading information in an efficient fashion.

The longitudinal components of the SOTs indicates a change in the magnetisation length that is intimately related to the spin accumulation direction.
This spin accumulation can add up or subtract from the magnetic moment of the ferromagnet depending on the direction of the electric field, which give rise to the unidirectional magnetoresistance (USMR)~\cite{Avci:2015jp}.
Although the longitudinal component is constant with the frequency when the magnetic moment points exactly along the spin accumulation direction, our results indicate that an enhanced AC-USMR may be obtained by rotating the magnetisation by a few degrees and close to the resonance frequency. 
This effect can be used in conjunction with pulsed perturbations as a mean to differentiate magnetic states.

Finally, the findings presented in this work offer the possibility of using frequency as a mean to manipulate the spin-orbit torques.
This opens a new road towards new dynamical control of the magnetisation dynamics for future spintronic-based devices based, for instance, on antiferromagnetic materials~\cite{Bhattacharjee:2018es,Olejnik:2018bk} or intrinsically 2D magnets~\cite{Gong:2019cd}.

\begin{methods}

\subsection{Electronic structure calculations}

We employ a multi-orbital tight-binding hamiltonian with nine orbitals (four for the $sp$ block and five for the $d$) and two spin channels per layer for a realistic description of the electronic structure of one ferromagnetic layer on top of four heavy metal layers for both Fe/W(110) and Co/Pt(001).
Density functional (DFT) calculations based on the real-space linear-muffin tin orbitals method within the atomic sphere approximation (RS-LMTO-ASA)~\cite{Andersen:1984ir,Peduto:1991hn,FrotaPessoa:1992kt} are used to provide the hopping matrix elements.
The ground-state hamiltonian includes an effective intra-atomic Coulomb interaction $U_i$ acting exclusively on the d orbitals, which is treated within mean field theory.
An on-site spin-orbit interaction $H_\text{SOI} = \sum_i \lambda_i\VEC{L}_i\cdot\VEC{S}_i$ (where $\lambda_i$ is also obtained from the RS-LMTO-ASA calculations) is also taken into account for the $d$ block of the hamiltonian.
We then self-consistently compute the ground-state magnetisation adjusting the center of the $d$-orbital levels to reproduce the DFT occupations while fixing the Fermi energy~\cite{bmcm}.
We use a constant broadening of the energy levels of $\Gamma=\SI{68}{\milli\electronvolt}$.

\subsection{Response functions}

The torque and magnetic responses of the system are obtained in the framework of linear response theory based on Kubo formalism~\cite{Kubo:1957cl}, employing the random phase approximation.
More details regarding the method used in this work is given in the Supplementary Note 1 and in Ref.~\citenum{Guimaraes:2017kk}.

\end{methods}

\noindent \textbf{Code availability} The tight-binding code that supports the findings of this study is available from the corresponding author on request.

\noindent \textbf{Data availability} The data that support the findings of this study are available from the corresponding authors on request.

\begin{addendum}

\item We are very grateful to R. B. Muniz and A. T. Costa for fruitful discussions and to A. B. Klautau for providing the tight-binding parameters. 
We also acknowledge the computing time on JURECA and JUQUEEN supercomputers at J\"ulich Supercomputing Centre. 
This work is supported by the European Research Council (ERC) under the European Union's Horizon 2020 research and innovation programme (ERC-consolidator grant 681405 — DYNASORE).

\item[Author contributions] F.S.M.G. performed the calculations and wrote the manuscript. F.S.M.G., J.B., M.d.S.D., and S.L. analyzed and discussed the results. All authors reviewed the manuscript. 

\item[Competing Interests] The authors declare no competing interests.

\item[Correspondence] Correspondence and requests for materials should be addressed to F.S.M.G. (email: f.guimaraes@fz-juelich.de) or to S.L. (email: s.lounis@fz-juelich.de).

\end{addendum}

\section*{References}

\bibliographystyle{naturemag}

\end{document}